\documentclass[a4paper,11pt]{article}

\usepackage[intlimits]{amsmath}

\usepackage{amsthm}

\usepackage{amsfonts}

\usepackage{latexsym}

\usepackage{epsfig}

\newcommand{\hs}{\hspace{1cm}}
\newcommand{\beq}{\begin{equation}}
\newcommand{\eeq}{\end{equation}} 
\newcommand{\eh}{\frac{1}{2}}

\newcommand{\R}{\mathbb{R}}       

\newcommand{\C}{\mathbb{C}}

\newcommand{\alf}{\alpha}

\begin{document}

\begin{titlepage}


\vskip 2.0cm

\begin{center}

{\large\bf D-Brane Scattering of $N=2$ Strings}\\

\vskip 1.5cm
{\large $\mbox{Klaus J\"unemann}^{a}\mbox{
 and Bernd Spendig}^{b}$}

\vskip 0.5cm
$ ^a$
{E-mail: K.Junemann@gmx.net}\\

$^b$
{\it Institut f\"ur Theoretische Physik, Universit\"at Hannover}\\
{\it Appelstra\ss{}e 2, 30167 Hannover, Germany}\\
{E-mail: spendig@itp.uni-hannover.de}\\

\end{center}
\vskip 2.5cm

\begin{abstract}
The amplitudes for emission and scattering of $N=2$ strings off D-branes 
are calculated. We consider in detail the amplitudes $\langle cc\rangle$ and
 $\langle occ\rangle$ for the different types of D-branes. For some D-branes 
we find massive poles in the scattering spectrum that are absent in the 
ordinary $N=2$ spectrum. 
\end{abstract}

\vfill

\textwidth 6.5truein
\hrule width 5.cm
\vskip.1in

\end{titlepage}

\section{Introduction}

 It has become obvious in the last couple of years that D-branes are of 
utmost importance for our understanding of $N=1$ string theory. In his 
pathbreaking paper \cite{polch}, Polchinski showed that $p$-dimensional
 extended objects -- the D-p-branes -- are the the long sought carriers
 of the Ramond-Ramond charges in $N=1$ string theory.  
In the context of scattering amplitude calculations the most important 
property of D-branes is that their quantum 
fluctuations are described by open strings moving on the brane and therefore
are under good control at weak coupling (for reviews on D-branes see 
e.g. \cite{bachas, JP} and literature cited within).
This allows to calculate amplitudes for emission and scattering of closed 
fundamental strings from D-branes by "pre-revolutionary" methods that have
 been invented more than a decade ago and are well understood \cite{FMS}. 
For the $N=1$  string these
computations have been performed e.g. in refs. \cite{KT,GHKM,GM,HK1,HK2} and 
considerably contributed to the understanding of D-brane physics. 
It is the purpose of this note to perform a similar analysis for $N=2$ strings.

 That this analysis has not been undertaken so far for the N=2 string finds 
its reason in the lack of Ramond-Ramond fields in the string spectrum. The 
$N=2$ superconformal algebra with $c=6$ serves as constraint algebra 
and is powerful enough to remove all string excitations from the spectrum
leaving the center of mass motion (which is not tachyonic but massless in 
this case) as the only physical degree of freedom \cite{OV}. This has 
immediate consequences for possible 
interactions. All n-point functions vanish \cite{hipp}, the only 
non-vanishing tree-level amplitude is the 3-point function. 
The corresponding field theory is self-dual gravity for closed strings and
 self-dual Yang-Mills theory for the open string sector. The critical 
dimension of the N=2 string is four but with ``unphysical'' signature (2,2), 
making it possible to identify the four real with two complex dimensions. 

Although lacking the necessary Ramond-Ramond fields, it is still possible to 
formally define D-branes in N=2 string theory by imposing Dirichlet boundary 
conditions in certain target space directions. The obvious question
is then whether the closed N=2 strings feel the presence of the branes. This 
note gives an answer to this question by performing a scattering analysis 
similar to the one undertaken in \cite{GM} for $N=1$ strings.

\section{Conventions}
We choose the flat target-space metric as $\eta^{\mu\nu} =$ diag$(-+-+)$.
It is advantageous to subsume the real $(2,2)$-vectors into complex 
$(1,1)$-vectors with metric $\eta^{\mu\nu}=diag(-,+)$.

In detail:
\begin{equation}
X^{\pm}=(X^{\pm 0},X^{\pm 1})=(X^0\pm iX^2, X^1\pm iX^3).
\end{equation}
The (2,2)- scalar product written in components reads 
\begin{equation*}
X_1\cdot X_2=\frac{1}{2}(X_1^+\cdot X_2^- + X_1^-\cdot X_2^+).
\end{equation*}
The (1,1)-scalar product is
\begin{equation*}
 X_1^+\cdot X_2^-=\frac{1}{2}\big(X_1^{++} X_2^{--} + X_1^{+-} X_2^{-+}\big),
\end{equation*}
where  $X^{\pm +}=X^{\pm 0}+X^{\pm 1}$ and $X^{\pm -}=-X^{\pm 0}+X^{\pm 1}$. 

Moreover, we introduce the matrices $J$ 
\begin{equation}
k^+ {\cdot} p^- = k {\cdot} p + i k {\cdot} J {\cdot}  p.  
\end{equation}
($J$ acts as a self-dual complex structure. It is $J_{02}=J_{13}=1$,
 $J_{13}=J_{02}=-1$, all other elements $=0$),
and  
\begin{equation}
D^{\mu\nu}=\mbox{diag}(D^{00},D^{11},D^{22},D^{33}).
\end{equation}

This matrix $D$ is related to the flat target space metric $\eta$ by a 
change of sign in the directions transverse to the D-brane.
Example: Let $x^2$ be the only direction transverse to the D-brane. Then 
$D=\mbox{diag}(-+++)$.

Emission and scattering off D-branes is conveniently calculated by evaluating 
 correlators between vertex operators on the upper half plane.
Open strings are  represented by holomorphic vertex operators  restricted to 
live on the real axis whereas closed string vertex operators factorize into
 holomorphic and antiholomorphic parts, 
\begin{equation}
V^{cl} (z, \bar z, p)  = :V(z, p/2): :V(\bar z, p/2):.
\end{equation}
Here z lies inside the upper half plane.
The sum of each picture number has to add up to $-2$ inside 
a non-vanishing scalar product. 
We will  use vertex operators in the $(-1,-1)$, $(-1,0)$ and $(0,-1)$ picture 
\cite{BKL}:
\begin{eqnarray}
V_{(-1,-1)}(k,z) &=& e^{-\varphi^- - \varphi^+} e^{ik{\cdot}X}(z), \nonumber \\
V_{(-1,0)}(k,z)  &=& k^+ {\cdot} \psi^- e^{-\varphi^-} e^{ik{\cdot}X}(z), 
\nonumber \\
V_{(0,-1)}(k,z)  &=& k^-  {\cdot} \psi^+ e^{-\varphi^+} e^{ik {\cdot}X} (z).
\end{eqnarray}

\section{The general calculations}

The separate propagators for holomorphic and antiholomorhic fields are
standard. However, due to the presence of a world sheet boundary, there are 
also non-vanishing correlation functions between holomorphic and 
antiholomorphic fields \cite{GM, IT1, IT2}:
\begin{eqnarray}
\langle X^{\mu} (z)  X^{\nu} (\bar w) \rangle &=& 
- D^{\mu\nu} \ln (z-\bar w), \nonumber \\
\langle \psi^{\mu} (z)  \psi^{\nu} (\bar w) \rangle &=& 
- \frac{D^{\mu\nu}}{z-\bar w}, \nonumber \\
\langle \varphi^{\pm} (z) \varphi^{\pm} (\bar w) \rangle &=& 
-  \ln (z-\bar w).
\end{eqnarray}
The example of the fermionic fields shows how this translates in the 
$\{\pm\}$-basis:

\begin{eqnarray}
k^- {\cdot} \psi^+ (z) p^- {\cdot} \psi^+ (\bar w) &\sim& - \frac{1}{z-\bar w}
\, k^- {\cdot}  (G_+ {\cdot} p)^+ , \nonumber \\
k^+ {\cdot} \psi^- (z) p^+ {\cdot} \psi^- (\bar w) &\sim& - \frac{1}{z-\bar w}
\,  k^+ {\cdot} (G_+ {\cdot} p)^-, \nonumber \\
k^+ {\cdot} \psi^- (z) p^-{\cdot} \psi^+ (\bar w) &\sim& - \frac{1}{z-\bar w} 
\, k^+ {\cdot} (G_- {\cdot} p)^-, \nonumber \\
k^- {\cdot} \psi^+ (z) p^+ {\cdot} \psi^- ( \bar w) &\sim& - \frac{1}{ z -\bar 
 w}  \, k^- {\cdot} (G_- {\cdot} p)^+.
\end{eqnarray}
with the definition $G_{\pm} = D \pm J \cdot D \cdot J$. 
The D-brane respects the complex structure in target space if $G_+=0$, i.e. 
$D = - J \cdot D \cdot J$. 

The new feature here (as compared e.g. to the mixed amplitudes) is that for 
Dirichlet boundary conditions in general one gets  poles in the 
operator product expansion between holomorphic and antiholomorphic fields 
 both having a  $+$ or $-$ index.

\subsection{$A_{cc}$}

 It was shown in  ref. \cite{GM} that 
the scattering amplitude of two $N=1$  closed strings off a D-brane can be
 obtained from the $N=1$ open string 4-point function by simply interchanging 
certain momenta. Thus the amplitude takes the form of an 
Euler-Beta-function of
 the Mandelstam variables that can be expanded as an infinite series of 
closed string 
poles in the $t$-channel or of open string poles in the $s$-channel and leads
 to the soft high energy behavior of the amplitude \cite{KT}. This 
result is intuitively clear since the interaction of closed strings with 
a D-brane is mediated by exchange of closed strings travelling between the 
passing closed string and the D-brane, or -- via world sheet duality --
by open strings moving along the brane. This argument should also be true in
 $N=2$ string theory. But it is difficult to imagine what a dual 
amplitude could look like in a theory with only a single degree of freedom. 
We therefore expect the scattering amplitude of a closed $N=2$ string off a 
D-brane to vanish\footnote{The D-instanton, of course, is an exception since
 in this case the 
$s$-channel point of view does not make sense and the scattering amplitude
is not required to be dual. In fact, as we will see the amplitude 
falls off as a power of $t$ being typical for a pointlike object.}.

The scattering amplitude of two closed strings off a D-brane for the $N=2$ 
string is given by the  integral of the correlation function of two closed 
string vertex operators with the correct quantum numbers over the upper
 half-plane 
$H^+$:
\begin{equation}\label{Acc0}
A_{cc}  (p_1, p_2)  \sim \int_{H^+}  d^2z \, d^2w \langle V_{(-1,0)} (z, p_1 /2)V_{(-1,0)} (\bar z, p_1 /2) V_{(0,-1)} 
(w, p_1 /2)
V_{(0,-1)} ( \bar w, p_1 /2) \rangle .
\end{equation}

Momentum conservation holds only in directions parallel to the brane:
\begin{equation}
\frac{p_1}{2} + \frac{D {\cdot} p_1}{2} + \frac{p_2}{2} + \frac{D {\cdot} p_2}
{2} = 0, \hs p_1^2 = p_2^2 = 0,
\end{equation}
where $p_1$ and $p_2$ denote the momenta of the incoming and outgoing strings,
 respectively.

In analogy to 4-particle scattering the amplitude can be parametrized by the
following Mandelstam variables:
\begin{equation}
s=(\frac{p_1}{2} + \frac{{D\cdot} p_1}{2})^2 , \hs
t = (\frac{p_1}{2} + \frac{p_2}{2})^2 , \hs
u = (\frac{p_1}{2}   + \frac{D{\cdot} p_2}{2})^2.
\end{equation}
Obviously $s$ is the momentum transfer along the brane and $t$ is the amount 
of momentum absorbed by the brane. As usual $s+t+u=0$.

 $SL(2,R)$ invariance of the correlation functions on the upper half plane 
allows us to fix three of the four variables of the vertex operators. For 
$A_{cc}$ we choose  $z  = iy$ ($y\in\R^+)$ and $w = i$.
The correct integration measure is $\int_0^1 dy (1-y^2)$. The resulting 
expression can be transformed into well known integral-representations of 
Euler-Beta-functions using the ``miracle''-substitution

\begin{equation} 
y = \frac{1-x^{1/2}}{1+x^{1/2}}.
\end{equation}
The final result is
\begin{equation}\label{Acc}
A_{cc} \sim  A   \frac{\Gamma(s - 1) \Gamma(t+1)}{\Gamma(s + t)} + B  
\frac{\Gamma(s) \Gamma(t)}{\Gamma(s + t)} - C  \frac{\Gamma(s) 
\Gamma(t+1)}{\Gamma(s + t + 1)}
\end{equation}
with
\begin{align*}
A&=p_1^+ {\cdot} (G_+ {\cdot} p_1)^-     p_2^- {\cdot}  (G_+ {\cdot} p_2)^+ ,
 \hs
B = 4(p_1^+ {\cdot} p_2^-)^2 , \\ 
C &=  (p_1^+ {\cdot} (G_- {\cdot} p_2)^-)^2.
\end{align*}

\subsection{$A_{ooc}$}
The amplitude for two open strings on the brane joining into
an outgoing closed string is
\begin{align*}
A_{ooc}   (k_1, k_2, p)  \sim & \int_{{\bf R}, x<y} dx \, dy  \int_{H^+}  
d^2z \\&\times 
 \langle V_{(-1,0)} (x, k_1)V_{(-1,0)} (y, k_2 ) V_{(0,-1)} (z, p /2)
V_{(0,-1)} ( \bar z, p /2) \rangle 
\end{align*}
$x$ and $y$ are integrated along the real axis in such a way that 
 $x$ is  always left of $y$.
The momenta $k_i$ of the open strings have to be parallel to the brane which
 implies $k_i = D {\cdot}k_i$. Momentum conservation in this case reads
\begin{equation*}
  k_1 + k_2 + \frac{p}{2} + \eh D {\cdot} p = 0
\end{equation*}
There is only a single kinematical variable $ s = k_1 {\cdot} k_2 = 
\frac{1}{4} p {\cdot} D {\cdot} p = - \eh p {\cdot} k_1 =
 -  \eh p {\cdot} k_2$.

To evaluate the amplitude $A_{ooc}$ we set $z = i$ and $x=-y$ ($x,y \in\R$) and
 use $2s=-u=-t $. 
The relevant integrals can all be evaluated with the formulae
\begin{equation*} 
\int_0^{\infty} dy \frac{y^a}{(1+y^2)^b} = \sqrt{\pi} 
\frac{\Gamma (b + \eh) \Gamma ( a+1) \Gamma (2b-a-1)}{\Gamma (2b) \Gamma 
(\eh 
a + 1) \Gamma (b - \eh a)}
\end{equation*}
and $\Gamma (a + \frac{1}{2} ) \Gamma (a) = \sqrt{\pi} 2^{1-2a} \Gamma (2a)$,
 resulting in
\begin{equation}\label{Aooc}
A_{ooc} \sim   (k_1^+ {\cdot} p^-) \, k_2^+ (G_-{\cdot} p)^- \cdot  
\frac{\Gamma ( 1-2t)}{\Gamma^2 ( 1 - t)}.
\end{equation}
This expression is completely analogous to that of the $N=1$ theory.

\section{Evaluating the general results for each D-brane-type}

In this section the above amplitudes are explicitly analyzed for each type 
of D-brane. Due to the special signature $(2,2)$ of our space-time we denote 
the D-branes by $p+q$, where $p$ and $q$ are the number of spatial and time 
directions, respectively, in which the D-brane lives. 

\subsection{The $(2+2)$ brane}
\subsubsection{$A_{cc}$}
In this case the brane fills all of space-time and we have ordinary interaction
between open and closed strings that has been considered in \cite{NM, SP}.
$A_{cc}$ has the interpretation of the lowest order quantum 
 correction to  closed 
string propagation.
Momentum conservation holds in all directions for closed string
 scattering off the $2+2$ brane. We cannot use our result, though, since by 
fixing three real parameters before integrating we did not divide out
 the volume of the whole symmetry group, which is, as we are dealing with a 
closed string topology $SL(2,\C)$ rather than $SL(2,\R)$. Naive application
 of our result (\ref{Acc}) would lead to $A_{cc}=0$, while the real amplitude 
is known to be constant.

\subsubsection{$A_{ooc}$}
For the process of joining of two open strings into a 
closed string momentum
conservation implies that $p{\cdot} k_1 = p{\cdot} k_2 = k_1{\cdot} k_2 = 0$.
Since $s= \frac{1}{4} p^2 = 0$ the amplitude (\ref{Aooc}) reduces to 
\begin{equation}
A_{ooc} \sim (k_1^+ {\cdot} k_2^-)^2
\end{equation}
coinciding with the well-known result \cite{NM}. 

\subsection{The $(1+2)$ brane}
\subsubsection{$A_{cc}$}
The $1+2$ brane divides space-time into two halves and is analogous to the 
$8$-brane in $N=1$ string theory.
There is only one transverse dimension which we choose to be the third. 
Momentum conservation  together with the mass-shell condition 
fixes the momenta in the closed string scattering process almost entirely. 

There are two cases; either the uninteresting case of no scattering at all, 
i.e. $p_1 = - p_2$, or the case 
\begin{equation}
p_1^0 = - p_2^0 ,\hs p_1^1 = - p_2^1 , \hs p_1^2 = - p_2^2, \hs p_1^3 = p_2^3. 
\end{equation}
The Mandelstam variables are $s = - t = \eh (p_1^3)^2$ and  $u=0$. 
What one finds from  eq.  (\ref{Acc}) is that  the first two terms vanish 
because the denominator diverges at $u=0$. The third term reduces to 
\begin{equation}
A_{cc} \sim -(p_1^+(G_\cdot p_2)^-)^2\, \Gamma(s) \Gamma(1-s) =-4[(p_1^0)^2  
+ (p_1^2)^2]\Gamma(s)\Gamma(1-s).
\end{equation}
Now using $\Gamma(s)\Gamma(1-s)=\frac{\pi}{\sin(\pi s)}$
we see that this expression has infinitely many simple poles
\begin{equation}
A_{cc} \sim  \frac{\big [ (p_1^0)^2  + (p_1^2)^2\big ]^2 }{\sin (\pi s )}.  
\end{equation}

\subsubsection{$A_{ooc}$}

Again we demand Dirichlet boundary conditions in the $3$-direction. 
$G_-=\mbox{diag}(0,-2,0,-2)$. The kinematics read $k_1^3=k_2^3=0$. We have
 to distinguish between two cases

a) $k_3^3=0$

Here $t=0$, thus we end up with a finite amplitude:
$$
A_{ooc}\sim k_{13}^+ k_2^+ \cdot(G_-\cdot k_3)^-.
$$
b) $k_3^3\neq 0$

We get 
$$
A_{ooc}\sim \frac{\Gamma(1-2t)}{\Gamma^2(1-t)} \sim \frac{\Gamma(1/2-t)}
{\Gamma(1-t)} \sim \cos(\pi\cdot(1/2-t))\Gamma(1/2-t)\Gamma(t).
$$
This amplitude has a tachyonic pole.

\subsection{The $(1+1)$ brane}
\subsubsection{$A_{cc}$}
For this kind of brane the matrix $D$ satisfies the relation
 $D =  J {\cdot} D {\cdot} J$ which implies $G_- = C = 0$ and $G_+ = 2 D$.
The closed string scattering amplitude becomes
\begin{equation}
A_{cc} \sim \bigg\{\frac{ p_1^+\cdot(D\cdot p_1)^-p_2^-\cdot(D\cdot p_2)^+}
{s-1} t  + (p_1^+p_2^-)^2  \bigg\} \frac{\Gamma(s) \Gamma(t)}{\Gamma(s + t )}.
\end{equation}
To further analyze the kinematical prefactor one recalls that in $2+2$ 
dimensions four momentum vectors with $\sum_1^4 k_i = k_i^2 = 0$ satisfy the 
relation \cite{hipp} 
\begin{equation}\label{hipp}
(k_1^+ {\cdot} k_3^-) (k_2^+ {\cdot} k_4^-) k_1 {\cdot} k_4 + (k_1^+ {\cdot} 
k_4^-) (k_2^+ {\cdot} k_3^-) k_1 {\cdot} k_3 = 0.
\end{equation}
It is this equation that is responsible for the vanishing of the 4-point
 function in open and closed $N=2$ string theory. 

Setting 
\begin{equation*}
k_1 = p_1 , \hs k_2 = D \cdot p_2, \hs k_3 = D {\cdot}p_1, \hs k_4 = p_2
\end{equation*}
and using the fact that  $(Dp_1)^- {\cdot} (Dp_2)^+ = p_1^+ {\cdot} p_2^-
$  for this particular form of $D$, one finds that
 $$(p_1^+\cdot(D\cdot p_1)^-p_2^-\cdot (D\cdot p_2)^+)t+(p_1^+p_2^-)^2 s = 0.
$$ What remains is
\begin{equation}
A_{cc}\sim   (p_1^+ {\cdot} p_2^-)^2 \frac{\Gamma(s-1) \Gamma(t)}
{\Gamma(s + t )}.
\end{equation}
In general this kinematical prefactor does not vanish and again the result 
has massive poles.

\subsubsection{$A_{ooc}$}

Since $G_-=0$ we have 
$$
A_{ooc}=0.
$$

\subsection{The $(0+2)$ brane}
\subsubsection{$A_{cc}$}
Two time dimensions with Dirichlet boundary conditions imply that $D = - J
 {\cdot} D {\cdot} J$ and therefore $G_+ = A = 0$ and $G_- = 2 D = 2
 diag(++++)$. The 
scattering amplitude is
\begin{equation}
A_{cc} \sim  [ (p_1^+\cdot p_2^-)^2u + (p_1 \cdot(Dp_2)^-)^2t ]  
\frac{\Gamma(s) \Gamma(t)}{\Gamma(s + t + 1)}.
\end{equation}

It is remarkable that the prefactor is related to  eq. (\ref{hipp}) by
 making the replacements 
\begin{equation}
k_1 \to p_1 , \hs k_2 \to D{\cdot} p_1, \hs k_3 \to p_2 , \hs k_4 \to 
D{\cdot}p_2 
\end{equation}
and using $(Dp_1)^+ {\cdot} (Dp_2)^- = p_1^+ {\cdot} p_2^-
$.
We therefore see that in this case where the brane does not break the complex
 structure in target space the scattering amplitude vanishes,
\begin{equation}
A_{cc} = 0.
\end{equation}

\subsubsection{$A_{ooc}$}

Open strings on this kind of brane are non-dynamical 
since the metric on the brane is euclidean such that the masslessness of the 
open strings implies the vanishing of their momentum.
The ${0+2}$ brane should therefore be thought of as a completely rigid object.
As one easily sees from inspection of the kinematical prefactor 
in eq.(\ref{Aooc}) the amplitude for closed string emission vanishes:

$$
A_{ooc}=0.
$$

\subsection{The $(0+1)$ brane}
\subsubsection{$A_{cc}$}
In this case all three terms in (\ref{Acc}) contribute to the  scattering
 amplitude $A_{cc}$ which can be rewritten as
\begin{equation}
A_{cc} \sim  \frac{1}{u(s-1)} \big ( A t u + B u (s-1) + C t (s-1) \big ) 
\frac{\Gamma(s) \Gamma(t)}{\Gamma(s + t )}.
\end{equation}
We checked that the  kinematical prefactor does not 
vanish for the values of $s$ and $t$ where the Beta-function has its poles.

\subsubsection{$A_{ooc}$}
Again this amplitude vanishes trivially.

\subsection{The $(0+0)$ brane / D-instanton}
\subsubsection{$A_{cc}$}
Dirichlet boundary conditions in all directions imply for the scattering
 process that there is no relation  between the momenta of the incoming and 
outgoing closed strings. Since $D=-\eta$ the Mandelstam variables and 
kinematical factors  become
\begin{align*}
&s=\frac{1}{4}p_1\cdot Dp_2=0,\hs t = -u = \frac{1}{4} p_1{\cdot} p_2, \\
& A = 0,  \hs B = C =  4 (p_1^+ {\cdot} 
p_2^-)^2 .
\end{align*}
The scattering amplitude in
 this case is
\begin{equation}\label{AccI}
A_{cc} \sim  (p_1^+ {\cdot} p_2^-)^2 \big(\frac{\Gamma(t)\Gamma(s)}
{\Gamma(t+s)}-\frac{\Gamma(t+1)\Gamma(s)}{\Gamma(t+s+1)}\big), 
\end{equation}
 leading to a simple $\frac{1}{t}$ pole at $t=0$. This pole can be obtained 
by either taking the $s\to 0$ limit in eq. (\ref{AccI}) 
or by recalculating the amplitude with $s=0$ from the very beginning. 

The single simple pole at $t=0$  
clearly is due to closed string exchange between the passing closed string
and the D-instanton.
The kinematical prefactor $ (p_1^+ {\cdot} p_2^-)^2 $ is precisely the 
3-point function of self dual gravity, as described by the Plebanski equation.
 From a field theory point of view the process should 
 be considered as
the scattering of gravitons off a pointlike (in space and time(s)) 
source, which can be identified with the D-instanton.

\subsubsection{$A_{ooc}$}

Needless to say that the amplitude for emission of a closed string off the
D-instanton vanishes.

\section{$A_{oooc}$}
For completeness we add our results we obtained calculating the amplitude
 for emission of a closed string from a D-brane on which three open strings
 interact.

We find that $A_{oooc}=0$ for all branes but the $1+2$-brane. In that case 
we are left with an integral of the type:
$$
\int dx \int dy \frac{1}{y(x+iy)}(x^2+y^2)^{\alf}((1-x)^2+y^2)^{\beta}
$$
($\alf,\beta \in \R$). So far we have not been able yet to solve this integral.

\section{Results}

We find that if the D-brane breaks the complex structure in target 
space additional correlation functions appear in the calculation which are 
absent for the usual Neumann boundary conditions. The result for the 
amplitude $A_{cc}$ is nevertheless an
 Euler Beta-function multiplied by a kinematical prefactor. 
A closer look at this kinematical factor shows that the amplitude vanishes 
 only  for the ${2+2}$  and the ${0+2}$ brane and has a single simple pole
at $t=0$  for the D-instanton which is due to closed string exchange.   
The scattering amplitudes of branes that break the complex structure 
in target space, i.e. the ${1+2}$, the ${1+1}$ and ${0+1}$ brane all have 
poles that do not correspond to states in the spectrum of the $N=2$ string. 

How do we interpret these results? In this paper we have considered the $N=2$
 string in its gauge fixed NSR formulation. 
Massive poles in the scattering spectrum seem to be inconsistent with this 
type of string. The inconsistency 
can be traced back to the fact the presence of these branes conflicts with 
$N=2$ world sheet supersymmetry. 
This is due to the fact that a fermion $\psi$ obeying Dirichlet boundary
 conditions cannot be in the same
 multiplet as a fermion obeying Neumann boundary conditions. This breaking 
of $N=2$ world sheet supersymmetry
 hence seems to leave no space for these type of branes in the gauge-fixed 
NSR formulation of the $N=2$ string.

This result is also consistent with T-duality. Recall that Neumann and
 Dirichlet boundary conditions are interchanged upon performing a T-duality 
transformation in a toroidally compactified space-time. However, for N=2 string
propagation only Ricci-flat K\"ahler manifolds with $(2,2)$ signature
are allowed. This leaves only the possibility to compactify one or both 
complex directions. Compactification of one or three real coordinates breaks
the complex structure and yields an illegal background.
Fortunately the three relevant branes,  namely the $2+2$- and  $0+2$-brane and 
the $D$-instanton, form a closed set under the action of T-duality in the
allowed backgrounds.

Apart from the 
$2+2$-brane the other relevant branes are non-dynamical since
  open strings attached to 
them have vanishing momentum\footnote{Note that dimensional reduction of 
abelian self-dual Yang-Mills theory to two timelike or zero dimensions yields
 a vanishing curvature two-form.}.
 This means that 
{\it dynamical} $D$-branes do not exist in the NSR formulation of $N=2$
 string theory in accordance with the absence of the corresponding 
differential forms and solutions of the classical equations of motion. We 
want to mention, though, that our formulation is not the only one that is 
able to describe the $N=2$ string. As was shown by 
Berkovitz and Vafa \cite{berko} and Siegel \cite{siegel} there exists as well
 a more general formulation of the $N=2$ string in terms of the so-called 
{\it topological} $N=4$ string. This formulation admits more degrees of 
freedom and hence there might be a way how the {\it forbidden} branes can be 
consistently incorporated in $N=2$ string theory. But so far no attempt in 
this direction has been made, leaving room for further work and speculations.

\noindent
{\bf\large Acknowledgements}\\[6pt]
We would like to thank O. Lechtenfeld for many fruitful discussions. 
B.S.\ thanks the Studienstiftung des deutschen Volkes for support.

\end{document}